# Impact of Cross-Listings on Chinese Stock Returns.

## A and N shares rate of return comparison.

Author: Kamilla Sabitova

Tutor: Han Li Yan

# Abstract

The paper examines the Chinese market reaction to the ADR issue by comparing returns and their stochastic variances of the Chinese firms cross-listed in the U.S. stock market.

First, it was implemented capital asset pricing model (CAPM) to determine expected returns A and N shares. The CAPM provided with a methodology to quantify risk and translate that risk into estimates of expected return on equity. Overall findings document that N shares of Chinese entities listed on U.S. market were greatly affected by economic turmoil during the period of World Financial Crises 2007-2008 than the A shares listed on the local market.

After, in order to test the hypothesis of beneficial cross-listing, it was implemented an event study method and the returns was modeled following GARCH process, which assumes homoscedasticity in residual returns. The results indicate a significant negative abnormal market return on an ADR listing date. The return volatilities after the listing date are compared to those before the listing. Four out of ten companies experienced increased volatility of local return after the cross-listing.

## Contents

# *Chapter I. Introduction*

## 1.1　　Research Background and Significance

Over the past two decades, an increasing number of firms, especially those from the emerging market countries, has cross-listed their shares on the major foreign stock exchanges around the world.What is the cross listing? Cross-listing is the listing of a company's common shares on a different exchange than its primary and original stock exchange. As the world economy becomes more interconnected, businesses around the world are looking for new ways to grow and enhance their competitive edge. There are many different ways for companies to increase their competitiveness. One way for company to grow is raising money on an overseas stock exchange. For example, in China, a listed company may have its shares in Hong Kong, in the United States, and in many different exchanges, in the world in order to raise new capital. In understanding the impact of cross-listing, I paid special attention to the situation in China.

　The cross-listing of Chinese companies has been growing since the resumption of the Chinese stock market in 1990 and has become a source of foreign capital inflows through international stock markets. In addition to foreign direct replenishment, China's economic investment began with the experience of cross-border listing of Chinese enterprises in 1980. With more and more Chinese companies listed on the international stock market, the research on China's cross-border listing is becoming a new focus of academic research. Some studies have investigated the price gap between the two listed Chinese securities (focusing on the Hong Kong-New York Double List); however there is a limited information whether the cross listing is a beneficial for emerging markets, and whether investment strategies could be developed from the price disparity phenomenon between dual-listed shares. There are several studies of the beneficial cross listing effect, documenting liquidity improvement and increase in home trading volume (Karolyi, 1996). However, some studies state that cross-listing shares do not create any value and that this strategy no longer appears to make sense. The research article of Richard Dobbs and Marc H. Goedhart (2008) presents that the trading volumes of the cross-listed shares of European companies in the United States account for less than 3 percent of these companies' total trading volumes; For Australian and Japanese companies, the percentage was even lower. However, companies from emerging markets may derive some benefit, but the evidence is not conclusive (McKinsey&Company: Strategy& Corporate Finance Article, 2008).

An Emerging markets, such as China, are often suffers by problems of low liquidity and insufficient regulation. All these problems are relevant to the Chinese stock market. The objective of this paper is to test the impact cross-listing on the emerging markets along with the comparison of the return of the dual-listed shares and analyzing which factors lead to the different returns of one company's shares listed on different stock markets. The study investigates the cross-listing effect via American Depository Receipts (ADRs) of Chinese firms. For the market indexes, SSE index and NYSE composite index were chosen for the local and overseas stock market indexes respectively. For more accurate and precise analysis, only those companies that trade on these markets were chosen.

### 1.2　Reason of Chinese Enterprises' Listing in the United States

According to the recently released World Bank data (2014), China overtook Japan as the world's second-largest economy. China's gross domestic product (GDP) has risen from 1.2 trillion U.S. dollars to nearly 500 trillion in the past decade. China's economy is becoming more strength and its impact on the number of cross-listed companies in the United States is also considerable. There is a significant amount of Chinese companies' listed in the United States stock market (Citibank, 2014).

When a foreign company decides to cross-list its shares in the United States market, the more stringent reporting requirements of the corporate entity must be met by American rules. This means that companies must register the U.S. Securities and Exchange Commission (SEC) and comply with U.S. securities laws and regulations. Strict reporting requirements may result in a choice of businesses, with U.S.-based exchanges such as NYSE or the Securities Dealers Association Automated Quotation (NASDAQ) cross-listing for reporting and compliance costs. These companies must allocate additional resources to meet the SEC and Sarbanes-Oxley testing and accounting requirements (SOX) reporting requirements. Despite these listing costs, companies already listed in the United States typically report lower capital costs, easier access to foreign capital markets, and increased shareholder base to improve stock liquidity and visibility and reputation (Doidge, 2004 ). The decision to cross-list must be a strategic one, ensuring that the expected return on cross-listing is greater than the cost.

**Table 1-1: Rank of the countries cross-listed their companies**

**in the United States (2010-2013):**

| Rank End 2010: | | Rank End 2013: | |
|---|---|---|---|
| 1. UK | 12% | China | 29% |
| 2. Brazil | 10% | UK | 11% |
| 3. Japan | 10% | Brazil | 9% |
| 4. Mexico | 7% | Japan | 6% |
| 5. China | 6% | Mexico | 5% |
| 6. Chile | 6% | Argentina | 4% |
| 7. Argentina | 5% | India | 3% |
| 8. Ireland | 5% | Chile | 3% |
| 9. Netherlands | 5% | Ireland | 3% |
| 10. France | 4% | Netherlands | 3% |

Looking at the cross-listing of companies over the period of 2010-2013, there are some interesting trends in the United States. Until 2010, cross-listed companies from the United Kingdom had the highest percentage in the United States. However, by the end of 2013, the Chinese claimed the top spot (Citibank, 2014). Cross-listing includes having shares trade in multiple time zones and in multiple currencies. This gives issuing companies more liquidity and a greater ability to raise capital. Listing shares by Chinese companies in Hong Kong or Singapore might be more viable as China's institutional environment and cultural practices may be closer to Hong Kong or Singapore, rather than to United States. Marcelo Bianconi and Liang Tan (2009), who observed data on public traded firms from six Asia-Pacific countries in 2003-2004, confirmed the hypothesis that cross-listing commands a premium on valuation; and growth opportunities further increase the premium. Authors found some evidence in favor of the US in the OLS and random effects models. However, they did not distinguish evidence in favor of US commanding higher premiums in the Asia-Pacific source countries. It was also mentioned that, unobserved factors that make cross-listing more likely to occur in the US are associated with lower valuations, hence cross-listing provides an opportunity for a firm to improve and achieve better market valuations.

### 1.3  Chinese Share Classes

Investing in China is tricky. There are now more than 20 China-focused ETFs (Exchange Traded Funds) to choose from. It seems that the cost rate and liquidity are not enough. Another big and fundamental factor that Chinese investors have to consider is China's share class.

Depending on the underlying ETF track, some funds are eligible to hold only one type of stock. This is important because ETFs in different classes of shares are qualified or non-compliant, and holding them can significantly affect the performance of the fund and ultimately determine the type of Chinese firms' portfolio.

There are three stock exchanges in the People's Republic of China (PRC): the Shanghai Stock Exchange, the Shenzhen Stock Exchange, and the Hong Kong Stock Exchange. Shares listed in Shanghai and Shenzhen Stock Exchanges are listed in RMB and re not entirely open to foreign investors due to capital account controls exercised by the Chinese mainland authorities. Both stock exchanges are non-profit organizations directly administered by the China Securities Regulatory Commission (CSRC). Shares listed in Shanghai Stock Exchange and Shenzhen Stock Exchange are called "*A Shares*" or "common shares", only domestic investors and selected foreign institutional investors are allowed to trade in A shares. Due to this restriction, most ETFs now enter the Chinese market through Chinese company stocks listed in Shanghai or Shenzhen in Hong Kong, the United States or special "B" stocks. These "investable" stocks include H shares, red chips, P-chips, B shares and N shares.

*B  shares:* Chinese companies incorporated in mainland China and traded in Shanghai, quoted in US dollars, or listed in Shenzhen and quoted in Hong Kong dollars (ownership of open foreign investment).

*H shares:* Chinese companies incorporated in the mainland and listed in Hong Kong.

*Red* **chips:** state-owned enterprises incorporated in mainland China (mainly in Hong Kong) and listed in Hong Kong.

*P-chip:* Chinese companies incorporated outside the non-state-owned mainland, most commonly in foreign (Cayman Islands, Bermuda, etc.) jurisdictions and listed in Hong Kong.

*N-shares:* Chinese companies incorporated outside the mainland are most commonly found in certain foreign jurisdictions and are listed on the New York Stock Exchange or the NASDAQ in the United States (H-shares and red chip adverse reactions are sometimes referred to as N - Unit).

The key to knowing which share class a fund is eligible to hold lies in knowing which index the fund tracks. FTSE, a leading index provider in the space, to date has considered P-chips to be Hong Kong companies, and has listed them in developed market Hong Kong indexes instead of their China index series. Therefore, ETFs that track FTSE's China indexes, such as FXI, are only eligible to hold H-shares and red chips, which limit their scope, as the indexers' classification methodology determines where that company sits in their global breakdown.

## 1.4　Literature Review

Since the financial crisis of 2007-2008, the global importance of emerging markets as China has increased. Some of the companies from the emerging world have succeeded in issuing large amounts of new equity through cross-listings in UK or US equity markets—something that might have been impossible at home. Karolyi (1996) provides evidence of the positive effect of globalization. He documents liquidity improvement, increase in total post-listing trading volume on average, and for many issues increase in home trading volume.

Finance theory suggests that increased market liquidity, market segmentation, disclosure and investor protection are advantages of cross-listing. Cross-listing enables companies to trade its shares in numerous time zones and multiple currencies. This increases the issuing company's liquidity and gives it more ability to raise capital. Foreign companies that cross-list in the United States of America do so through American depository receipts (ADR). However, Dobbs and Goedhart (2008) argues that institutional investors typically invest in stocks they find attractive, no matter where those stocks are listed and that cross-listing no longer appears to make sense.

One of the earliest studies of the cross-listing effect was done by Jayaraman, Shastri and Tandon(1993), who examined the impact of the listing of 95 American Depository Receipts (ADRs) from Japan, Germany, United Kingdom, Italy and positive abnormal returns to the underlying stock on the listing day, mainly driven by Japanese entities. Authors interpret this evidence as consistent with the existence of informed traders in the markets in which the ADRs and the underlying stocks trade. Mittoo (2003) examined a sample of Canadian companies listed in the United States between 1991 and 1998. The study showed that for a period of seven days around the event date stock prices showed significant growth with an average excess return of 0.68%. Korczak and Bohl (2005), who analyzed ADR' from 33 Central and European countries traded since 1995 to 2004, also found a significant (1%) positive response to the market, not just the first day but a cumulative abnormal return of 26% over a longer period of time (50 days, 200 days).

There were also conflicting findings whether cross-listing beneficial. For example, Howe & Madura [1990] studied firms from the US listing in Germany, France, Japan and Switzerland between 1969 and 1984. The results show that on average both the US and the host country Beta decreases significantly. The total risk, defined as the Standard deviation, is not affected. They conclude that a dual listing is an ineffective tool for reducing segmentation. More evidence were found by Varela & Lee [1993] who studied the cost of capital effects for 168 US companies listed on the London Stock Exchange. The cost of capital decreased slightly during the first sub period (-1984) while during the second sub period (1984- ) the decrease were greater. The reason for this was the higher listing costs and the "Big Bang" on the LSE 1986. The total risk for the companies was lowered due to the cross-listing. In addition, there was also the presence of studies recording mixing results. Barclay (1990) found that the volatility before and after cross-listing did not change. It contradicts with the results of Jayaraman and Makhija (1990), who found a significant increase in the variance of return on issuance after adverse reactions were found.

If we consider the share price action related to the analysis of the incident, such as placement of depositary receipts, for example, in an article Miller (1999) shows a rights issue that generates an average of positive returns on excess earnings over a three-day interval Capabilities have a positive impact: they have done 2.63 and 1.27% for companies who have placed first-tier and 144A ADRs, second and third tier companies - 1.83 and 3.23% respectively. Bessler ，Kaen，Kurman ，Zimmerman（2012）presented the study of 18 German companies cross-listed on the New York Stock Exchange and NASDAQ in 1990-2005, showing that the underlying companies did not receive significant benefits during their listing. On the contrast there were determined a positive effect of delisting (the average cumulative excess return was within three days of the announcement of the delisting 1.22%).

The theoretical basis of the capital asset pricing model, first developed by Sharp in 1963, was followed by an equilibrium model of the price of capital markets by Sharpe (1964), Linter (1965), and Mossin (1966). The systematic and nonsystematic risk refinement of the treatment of risk in capital asset pricing models was conceptualized by M. Markowitz in 1950.

Among the recent studies of beneficial effect of cross-listing on emerging markets, Smirnova (2004) by investigation the sample of Russian ADRs indicated that the conditions under which ADR introduction from emerging markets affect the underlying stock are not well clarified.

One of the first attempts to investigate China's stock market was by Sun Huibing (2008). He studied the impact of securities in Hong Kong and in Chinese mainland market, and distinguished a negative impact.

The conflicting results of previous works have questioned the efficiency of international cross-listing. It shows the importance of further investigation of the effect. The study intends to give some evidence of cross-listing related implications for China. Therefore, it is difficult to identify a clear trend in the market and to make a clear conclusion whether value of the listed company has a positive effect.

**1.5　Methods and Purpose of Dissertation**

The purpose of the dissertation is to present a qualitative analysis on the term of cross-listing. Theories and empirical results in the subject differ, depending not only on which stock exchanges are examined, but also on the listing period of the company. The previous research will be extensively examined. First, to see which class of shares has more beneficial return for investors from a cross-listing. Second, is to test the hypothesis of beneficial cross-listing effect on local stock returns. Furthermore, the analysis will investigate the following aspects:

- The comparison of stochastic volatility of different class of shares of the same company

- What is the the relationship between risk and expected return of one company's different class shares?
- What factors contributes to high risk and low expected return?
- What factors influence the listing location?
- Why does the return volatilities after the listing different to those before the listing?

In order to achieve the research goal and strive to make the research scientific and reasonable, this paper mainly uses the following two methods. In Chapter 2, I will explore the proceeds from the stock. By implementing Capital Asset Pricing Model (CAPM), I will able to explain the relationship between risk and expected return of A and N shares. In the capital asset pricing model, the risk associated with an asset is a measure of the overall risk associated with the market. In Chapter 3, I used an event study method to test the hypothesis of beneficial cross-listing effect. When using event study method, I used GARCH model the same as Smirnova (2004). I calculated the abnormal returns and the cumulative abnormal returns 15 days before and after the listing date in order to determine whether the cross-listing has a beneficial effect for the stocks from developing markets. Then I will compare the volatilities of stock returns 90 days before and after the ADRs listing date.

I used XLStat, MATLAB and Eviews software in this study.

## Chapter II. N and A Shares Rate of Return Comparison

### 2.1 Overview

The purpose of this chapter is to compare the rate of return of A and N shares and determine stocks' behavior of the same companies listed in American and Chinese stock markets. The cross-listing is a particularly common occasion for the companies that started in a small market but grew into a bigger market. I chose Chinese firms, that have a listings in both New York Stock Exchange and Shanghai Exchange, only ten Chinese companies suit this criteria.

**Table 2-1: Researched Companies**

| Company Name | A Code | N Code | Industry | Market Cap, US$ | Listing date on A stock market |
|---|---|---|---|---|---|
| Sinopec Shanghai Petrochemical Co. Ltd. | 600688 | SHI | Oil & Gas Producers | 5,35B | 08.11.1993 |
| Guangshen Railway | 601333 | GSH | Travel &Leisure | 4.31B | 13.5.1996 |
| China Petroleum & Chemical | 600028 | SNP | Oil & Gas Producers | 104,76B | 08.08.2001 |
| Huaneng Power International | 600011 | HNP | Electricity | 18,13B | 06.12.2001 |
| China Southern Airlines Co. Ltd | 600029 | ZHN | Travel & Leisure | 9,03B | 25.07.2003 |
| China Life Insurance | 601628 | LFC | Life Insurance | 141,79B | 09.01.2007 |
| China Eastern Airlines Co. Ltd | 600115 | CEA | Travel & Leisure | 11,61B | 11.15.1997 |
| Aluminum Corp. of China Ltd | 601600 | ACH | Factory | 8,68B | 30.4.2007 |
| Petro China | 601857 | PTR | Oil & Gas Producers | 240,43B | 5.11.2007 |
| China Unicom | 600050 | CHU | Mobile Telecom | 36.119B | 21.6.200 |

It was chosen an Event window for the past 14 years from January 2001 to December 2014, as the global market during this period has been greatly affected by the World Financial Crisis and the FED Tapering.

**2.2 Rate of Return and Stochastic Volatility Comparison**

    The first step was to calculate the Rate of Return and stochastic volatility of the stock market. Rate of Returns was calculated as follows:

$$R_t = \ln P_t - \ln P_{t-1} \quad (1)$$

    We know that prices of different financial assets, such as currencies and stocks, are constantly fluctuating, as traders buy and sell these assets. Changes in prices over the time arouse volatility. Volatility is a statistical measure of the dispersion of rate of returns for a given security or market index; the higher volatility, the riskier the security. Currency pairs are volatile and involve high risk, but there is also the opportunity to earn profits from Forex Traders. There is a strong relationship between volatility and market performance. As the volatility increases, the risk also increases and the return decreases. The risk is represented by the spread of the average return around it. In this chapter, I will measure volatility using standard deviation and determine how close the stock prices are around the average or moving average combination. By using historical data of NYA (for the index return of foreign market) and SSE (For the index return of domestic market) indexes data over the past 14 years, I am able to calculate the volatility. Because investors are very concerned with how well their investments are performing, knowing how to gauge such performance is important. Rate of return of the stock refers total income earned by the stock investment and the ratio of the original investment. Stock is favored by investors because of the benefits of buying stocks. Figure 2-1 shoes the return of A and N shares. The chart depicts that both A and N shares had almost the same rate of return performance in the period from 2002 to 2012. However, before 2002, the returns of both kinds of shares were in different directions. Since 2012, the rate of return of N-shares has risen, and the return of A-shares has dropped and suddenly rose in 2013 from negative to positive rate. A negative rate of return is a financial term that refers to a business that fails to earn profit on a specific period of time. It also can be referred to the value of capital investments such as loss on stocks, commodities or real estate. While investing in a new business as a whole is often negative in the first period of

establishing a business as a whole, a negative return does not necessarily indicate an unqualified business. In the stock market, there is a common situation when the most investments are made when the market fluctuate more for the ultimate control of public traded entities over the valuation of their stocks.

**Figure 2-1: A and N Shares Rate of Return**

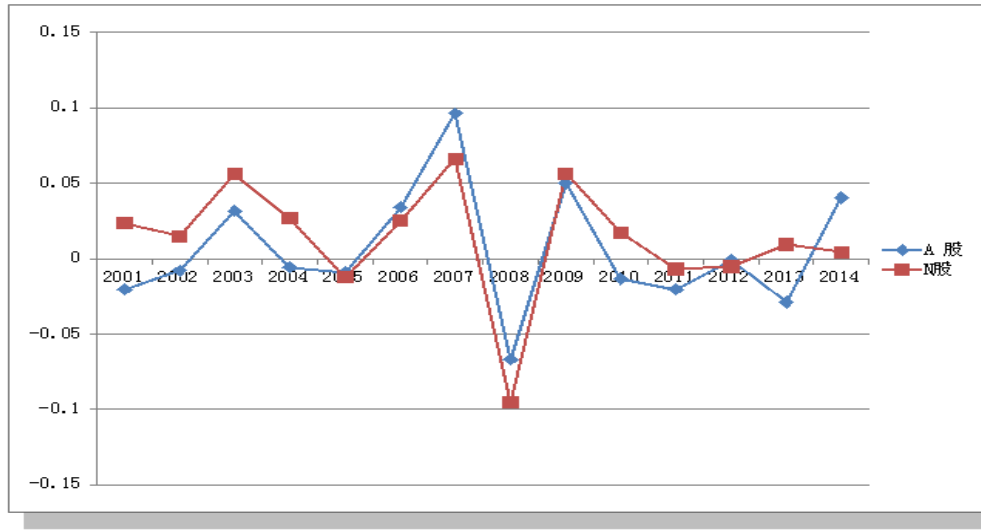

Then, by using the standard deviation term, I measured volatility and determined how close the stock prices of the nine given Chinese firms are around the mean. Figures 2-2 and 2-3 represent the volatility of A and N shares:

**Figure 2-2: A-Shares Volatility**

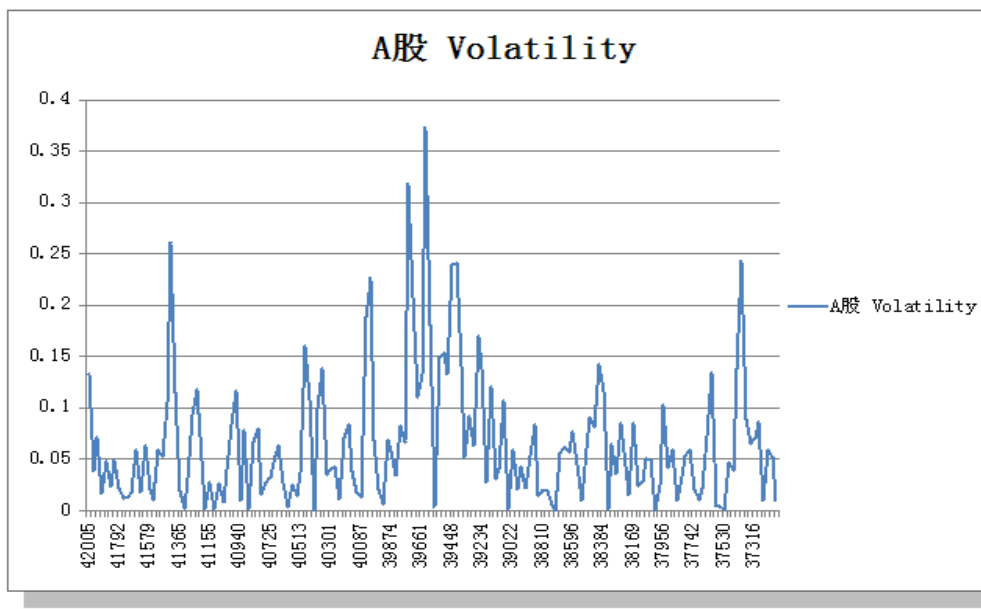

**Figure 2-3: N-Shares Volatility**

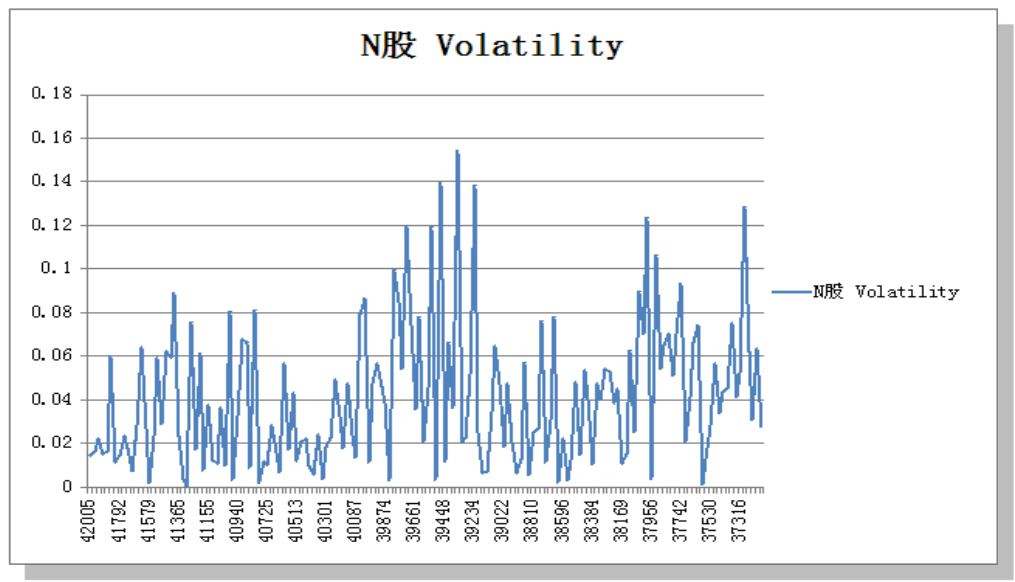

From the Figures 2-2 and 2-3, it is clearly seen that both shares experienced high volatility during the High-Tech Bubble period and World Financial Crisis 2007-2008. On May 22, 2013, when Federal Reserve Chairman Ben Bernanke stated in testimony before Congress that FED may taper- or reduce- the size of the bond-buying program known as a quantitative easing. This program, was designed to stimulate the economy, has served the secondary of supporting financial market performance recently. From the Figure 2-2, we can see that the volatility of both shares suddenly starts to rise in 2013. In the recovery that has followed the 2008 financial crisis, both stocks and bonds have produced outstanding returns despite economic growth that is well below historical norms. The general consensus, which is likely accurate, is that Fed policy is the reason for this disconnect. In 2013, the markets would start to perform more in line with economic fundamentals – which in this case, meant weaker performance. However, after the market stabilized in the second half of 2013, investors gradually became more comfortable with the idea of quantitative easing cuts. This shows why the volatility of A-shares fluctuated suddenly in 2013. The volatility of A-shares is relatively stable, which referred as a less risky as N-shares.

## 2.3 Expected Return Comparison

The second step is the comparison of Expected Rate of Return of A and N shares. Expected Return is the amount of profit or loss an investor anticipates on an investment that has various known or expected rates of return. Comparison of the Expected Return of A and N- shares provides an indication of both shares performance in foreign and local markets during the given period of time. By determining the measure of volatility, or systematic risk, β coefficient, I will able to calculate the expected return of an asset and expected market returns. Beta is the key factor used in the Capital Asset Pricing Model (CAPM), a model that measures the return of a stock. The theoretical basis of the capital asset pricing model, originally developed by Sharp in 1963, was followed by an equilibrium model of the price of capital markets by Sharpe (1964), Linter (1965), and Mossin (1966). By using CAPM I will able to test market integration, investigating investment barriers and analyzing the behavior of market indices. For the data itself, the monthly figures, from January 2001 to December 2014, are calculated as average high and low prices. CAPM needs to leverage the market portfolio. The market portfolio is represented by market indices, for A-shares and N-shares SSE indices and NYA indices, respectively, were used. The yield of a stock is calculated using Equation (1). The following regression equation is used for test evaluation:

$$R_{it} = \alpha_i + \beta_i Rm_t + \varepsilon_t \quad (2),$$

Where $R_{it}$ is the return rate of the stock at time t, and $Rm_t$ is the market rate of return at time t. By estimating β, I conducted the ordinary least squares (OLS) method. One of the assumptions underlying OLS estimation is that the errors be uncorrelated. Since the autocorrelation of the errors is a violation of regression analysis, Durbin Watson Test was used to test autocorrelation. DW-test measures the linear relationship between residuals adjacent to the regression model. If there is no sequence correlation, test statistics will be about 2. If there is a positive sequence correlation, the test statistic will be below 2, and if there is a negative correlation, the statistic will be between 2 and 4. Table 2-2 depicts DW-test results:

**Table 2-2: DW-Test Results:**

| 股票种类 (Type of Shares) | DW-Test Statistics |
|---|---|
| A 股 (A-shares) | 2.007 |
| N 股(N-shares) | 1.96 |

The result of DW test excludes the presence of autocorrelation, what means that error terms are not serially correlated and there are no issues will arise while using OLS.

The alpha and beta are both metrics relating to the level of risk or volatility experienced in a particular security. While the alpha provides a measurement in regards to the assets performance, the beta specifies the level of risk present when compared to the capital asset pricing model (CAPM). Calculated using a form of regression analysis, the beta is a measure of the asset's ability to respond to market fluctuations. Volatility of stocks and systemic risk can be judged by calculating β. In the capital asset pricing model, the risk associated with an asset is a measure of the overall risk associated with the market. This is expressed as β of the stock. A β of less than 1 asset will show that the average change in returns is less extreme than the overall market, β will be greater than 1 and the performance will fluctuate more than the overall market. The higher the β, the steeper the value of the investment can be expected to fluctuate relative to the market index. Funds with β greater than 1 are considered more volatile than the market; less than 1 means less volatility. Table 2-3 shows the OLS results, α and β estimation results:

**Table 2-3: OLS Results**

| 股票种类 | 估计 α | Standard err | t-stat | 估计 β | Standart err | t-stat |
|---|---|---|---|---|---|---|
| **A 股** | **0,004290** | **0.003669** | **1.169531** | **0.923423** | **0.166425** | **5.548\*** |
| **N 股** | **0.001549** | **0.007439** | **0.208206** | **2.642174** | **0.230962** | **11.44\*\*** |

Roughly speaking, the A shares fluctuate around 0.923, what means that A-shares theoretically less risky; for the N shares beta is 2,64, offering the possibility of a higher rate of return, but also posing more risk. A-shares' β below 1, what means that the price is expected to be more stable and less volatile.

Since I have estimated β and α for both types of stocks, I can find their expected return. The Expected Return is calculated as follows:

$$E(R) = R_f + \beta \, (R_{market} - R_f) \ (3) \ ,$$

Where $R_f$ is a risk-free rate. The risk-free rate represents the interest an investor would expect from an absolutely risk-free investment over a specified period of time. In theory, the risk-free rate is the minimum return an investor expects for any investment because he will not accept additional risk unless the potential rate of return is greater than the risk-free rate.

In practice, however, the risk-free rate does not exist because even the safest investments carry a very small amount of risk. Thus, the interest rate of China Government bond yield were used for A-shares (data source http://www.tradingeconomics.com) and US government bond yield were used for N-shares (data source http://www.bloomberg.com). Table 2-4 shows the Expected Return on A and N shares:

**Table 2-4: Expected Return Results:**

| | |
|---|---|
| **A 股** | **0,193505289** |
| **N 股** | **-3,473188235** |

Table 2-4 documents that the expected rate of N shares is negative, and the A shares are positive.

Several factors can cause the presence of negative rate of return, such as poor performance of a company, inflation and turmoil of the entire company. However in this case, negative rate of return is not cause by problems related to companies' performance. Turmoil of the economy as a whole can result in negative rate of return. Because of the financial crisis 2007-2008, during which the broader market lost over 50% of its value, the U.S. market was under great pressure and the stock market experienced a high volatility. Therefore listing of Chinese firms on American stock market had a different rate of return with listing in Chinese stock market. Overall findings document that the cross-listing was beneficial for emerging markets, as China, because investors was appealed to invest in less risky environment.

## *Chapter III. The Impact of Cross-Listing on Chinese Stock Returns*

### 3.1 Overview

Most foreign companies that cross-list in the U.S. markets do so via American Depository Receipts. As documented by JP Morgan (2003), the history of the first depository receipt program goes back to 1920s. The UK-based retail company "Selfridge Provincial Stores Ltd." decided to expand its shareholding base and aimed to start selling its shares to the US investors. American Depository Receipts, invented in 1927, gave American investors an opportunity to purchase stock of foreign firms without concern of settlement delays. Over the past two decades, one of the most subsets of the markets has been the rapidly rise Chinese ADR market. According to the CitiBank data, Chinese companies' cross-listed in the U.S. stock market overtook United Kingdom from the period of 2010-2013. China uses this fact to actively expand its financial and economic influence around the globe, be it by promoting the international use of the Renminbi (RMB), by increasing China's IMF and World Bank quota and voting rights, or by acquiring or trying to acquire strategic foreign assets and natural resources. Following by the findings in Chapter II, expected returns of investment of Chinese Companies in local market has been positive during the last ten years. Listing of Chinese firms on American stock market had a different rate of return with listing in Chinese stock market; shares of Chinese entities listed in the U.S. market was affected by the economic turmoil, therefore presented a negative expected return.

### 3.2 Event Study

Measuring the effects of economic event is frequently asked by economists. The usefulness of such study stems from the fact that in the rational market, the influence of an event can be quickly reflected in asset prices. As a result, the economic impact of an event can be measured using the asset prices observed over a relatively short period of time. In contrast, direct measure of the economic impact may require several or even a few years ob observed data. There are 95 Chinese Companies issuing ADRs listed on the US Exchanges and 200 Chinese firms trading on the US OTC Markets    (http://topforeignstocks.com/foreign-adrs-list/the-full-list-of-chinese-adrs/). The sample that suits my scope consists of 10 companies domiciled in China and issuing ADR, listed in New York Stock Exchange and Shanghai Stock Exchange.

An event study method is used to measure changes in share value around the listing date. Local returns are computed as follows:

$$R_i = \ln P_t - \ln P_{t-1} (1),$$

Where $P_t$ is daily closing price.

In this chapter, to measure abnormal return I estimated a market model for each researched firm using local stock returns denominated in US dollars. The event window was chosen with same techniques of SUN HUIBING (2007) work, with the listing date defined as 0, the market coefficients are estimated in the pre-listing period: day -105 to day -15.

According to Campbell (1997), estimation of excluding 51 days around listing date is typical and the event window will not be overlapped. Such model design provides the estimators for the parameters of the normal return that are not influenced by the event-related returns. So, the so-called selection bias may exist if an estimation period is chosen very close to the launch date. It leads to a bias towards the expected return on the stock before the market. The periods of non-overlapping ADRs listed on day zero are shown in Figure 3-1:

**Figure 3-1：Event Window**

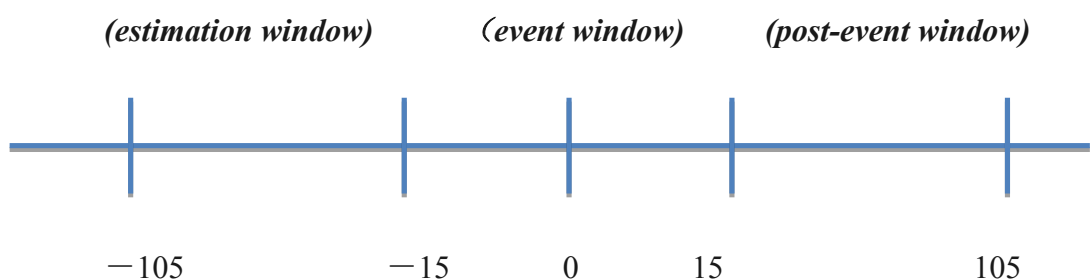

**Table 3-1: Researched companies:**

| Company Name | A Code | N Code | Industry | Market Cap, US$ | Listing date on A stock market |
|---|---|---|---|---|---|
| Sinopec Shanghai Petrochemical Co. Ltd. | 600688 | SHI | Oil & Gas Producers | 5,35B | 08.11.1993 |
| Guangshen Railway | 601333 | GSH | Travel &Leisure | 4.31B | 13.5.1996 |
| China Petroleum & Chemical | 600028 | SNP | Oil & Gas Producers | 104,76B | 08.08.2001 |
| Huaneng Power International | 600011 | HNP | Electricity | 18,13B | 06.12.2001 |
| China Southern Airlines Co. Ltd | 600029 | ZHN | Travel & Leisure | 9,03B | 25.07.2003 |
| China Life Insurance | 601628 | LFC | Life Insurance | 141,79B | 09.01.2007 |
| China Eastern Airlines Co. Ltd | 600115 | CEA | Travel & Leisure | 11,61B | 11.15.1997 |
| Aluminum Corp. of China Ltd | 601600 | ACH | Factory | 8,68B | 30.4.2007 |
| Petro China | 601857 | PTR | Oil & Gas Producers | 240,43B | 5.11.2007 |
| China Unicom | 600050 | CHU | Mobile Telecom | 36.119B | 21.6.200 |

Abnormal returns in the event window are determined by calculating errors from the market model. Coefficients from the pre-listing model are used to calculate abnormal returns from day -15 to day 15. Abnormal returns are then averaged across 1-researched companies (Average Abnormal Returns) and across time (Cumulative Abnormal Returns)

Stock returns are model using the GARCH model (Smirnova 2004), instead of OLS method used in many previous studies. The existence of residual income in the case of heteroskedasticity leads to inefficient estimation of many parameters of the asset pricing model. There are also many studies investigating unconditional heteroskedasticity models, but finding problems that are more often encountered with conditional heteroskedasticity. In order to consider the market model of residuals with conditional heteroskedasticity, such as Generalized Autoregressive Conditional Heteroskedasticity (GARCH) model, another method of estimation is implemented. It also shows that this technique improves the efficiency of the coefficients (Akgiray, 1989). According to Dielbold (1992), p and q are market models for introducing firm errors into the GARCH process, i is the following,

$$R_t^i = \beta^i X_t^i + \varepsilon_t^i, \quad \varepsilon_t^i \mid \Omega_{t-1}^i \approx N(0, h_t^i) \quad (2),$$

where $h_t^i = \alpha_0^i + \sum_j \alpha_j^i (\varepsilon_{t-j}^i)^2 + ... + \sum_k \gamma_k^i (h_{t-k}^i)$, $j = 1,...,q, k = 1,...,p$ the conditional variance of the error term，

$R_t^i$ is the return on the stock of firm $i$ at time $t$ (in USD)，

$\Omega_{t-1}^i$ is the information set，

$X_t^i$ is the row vector contains intercept, the local market index return and the US market index return.。 For the domestic index return I chose the SSE index, as it is the oldest indicator of growth in China's capital market; NYSE composite index return is used as a proxy for US market return, as it is a float-adjusted market-capitalization weighted index which includes all common stocks listed on the NYSE, including ADRs

To examine the heteroskedasticity I implemented the hettest in Eviews to determine the benefit of the Breusch-Godfrey test regression. I found that most of the securities return showed heteroskedasticity.

Then I need to test whether the number of lag errors (q) and the conditional variance depend on the preceding variance (p). I know that the most common application in practice is the GARCH (1,1) model. First, to establish the GARCH model, I estimate the different lag numbers for a security and choose a likelihood function and a statistically significant coefficient at the maximum. After that, I got the estimated coefficient of the market model, and I can calculate the stock abnormal return (*AAR*) according to the following formula at time t:

$$AR_{it} = R_{it} - (X_t^i \beta^i) \quad (3),$$

where $\beta^i$ is the vector of the estimated intercept and the coefficients for local market and US market indexes from the market model (2)

The daily abnormal returns are then averaged across *N* securities, weighted by companies' size, on day *t* to compute the average abnormal return（*N* = 10）：

$$AR_t = \sum_{i=1}^{n} AR_{it} w_i \quad (4),$$

$$w_i = \frac{Mcap_i}{\sum_{i=1}^{n} Mcap_i} \quad (5),$$

where，

Cumulative average abnormal returns were calculated as follows：

$$CAR_{ab} = \sum_{t=a}^{b} AR_t \quad (6),$$

Where, *a* and *b* are the beginning and the end of the estimated window.

To test the statistical significance of the abnormal return of the method, I used the method proposed by Dodd and Warner (1983). For each stock I, the daily excess return rate is normalized by the square root of its daily variance of t by its estimated variance:

$$StAR_{it} = \frac{AR_{it}}{s_{it}} \quad (7),$$

where，
$$s_{it} = \sqrt{\frac{s_i^2\left[1 + \frac{1}{L} + \left(R_{mt} - \overline{R}_m\right)\left(R_{ust} - \overline{R}_{us}\right)\right]}{\sum_{k=1}^{L}\left(R_{mk} - \overline{R}_m\right)\left(R_{usk} - \overline{R}_{us}\right)}} \quad (8),$$

$s_i$ is the estimated residual variance from the market model regression for security $i$, $R_{mt}$ is the SSE return on day $t$, $R_{ust}$ is the NYSE return on day $t$, $\overline{R}_{US}$ is the mean return on the NYSE index over the $L$ days（$L$＝90）.

Z statistics is calculated as follows:

$$Z_t = \sum_{i=1}^{n} StAR_{it}\sqrt{\frac{1}{n}} \quad (9),$$

where $n$ is a number of researched firms ($n$=9)

Average cumulative standardization statistics is calculated as follows:

$$CZab = \sum_{t=a}^{b} Z_t \sqrt{\frac{1}{b-a+1}} \quad （10）$$

Standardization is used in the normalization process in order to be able to accumulate the overall distribution of residuals compared to normal cell roots. Using standardization allows averaging accumulated residuals and comparing two variables that are from different normal distributions.

### 3.3 Event Method Results and Analysis

On the listing day, day 0, the local market shows a negative significant return on average of 4.63%. Table 2-2 shows the daily average abnormal returns ($AR_{it}$), and the cumulative abnormal returns ($CAR_t$) for day -15 through day 15 around the listing date. The results agrees with Lau, Dlitz and Apilado (1994)m who found the presence of negative valuation of US companies, that became public entities, on the first trading day . The authors also found a significant negative cumulative average abnormality within a trading day of the interval [-5, +3]. Javaraman finds the result of a positive daily excess return (1993) of ADR to US-listed foreign companies. This is in contradiction with the result of his (1993) finding, who found that the ADR of foreign entities listed on American market had a positive daily abnormal return.

*Table 3-2 : Average Abnormal Returns (ARit), Cumulative Abnormal Returns (CARt) and Z-statistics (\*significant at level 0.05)*

| Days | $AR_{it}$ | $CAR_t$ | $Z_t$ |
|------|-----------|---------|-------|
| -15 | 0,0501 | -0,101 | -0,027842922 |
| -14 | -0,0299 | -0,151 | 0,015722873 |
| -13 | 0,0281* | -0,121 | -0,048711739 |
| -12 | -0,0362 | -0,149 | -0,017419035 |
| -11 | 0,0111 | -0,113 | -0,003886976 |
| -9 | -0,0308 | -0,124 | 0,010586266 |
| -8 | -0,0337 | -0,0929 | 0,023958863 |
| -6 | 0,0255 | -0,0592 | -0,031080363 |
| -5 | -0,0554 | -0,0847 | 0,000994693 |
| -2 | -0,00814 | -0,0293 | 0,02089594 |
| -1 | 0,0368 | -0,0212 | -0,027842922 |
| 0 | -0,0463* | -0,058 | 0,048228318 |
| 1 | -0,0156* | -0,0116 | 0,054311869 |
| 2 | -0,0327 | 0,00402 | -0,016106196 |
| 5 | 0,037 | 0,0367 | -0,023510819 |
| 6 | -0,0101 | -0,00024 | 0,03034005 |
| 8 | 0,0108 | 0,00987 | 0,031483831 |
| 9 | 0,0000747* | -0,00097 | -0,040595025 |
| 11 | -0,00929 | -0,00104 | -0,031427064 |
| 13 | 0,016 | 0,00825 | 0,020798922 |
| 14 | -0,00215 | -0,00771 | 0,007901331 |
| 15 | -0,00556 | -0,00556 | 0,00435158 |

Most of the cumulative abnormal returns are negative and statistically insignificant. Figure 202 depicts cumulative abnormal returns in the given event window:

The finding proves the hypothesis of beneficial cross-listing effect for the stocks from developing markets. However, most of results are statistically insignificant, what suggests that introduction of ADRs in US market does not have a significant impact on the underlying Chinese stocks and consistent with Martell's, Rodriguez' and Webb's research, examining the risk and return of several Latin American equities

**Figure 3-2: Cumulative abnormal returns from day –15 before to day +15 after the listing of an American Depositary Receipt program. The daily abnormal returns are averaged across firms and then cumulated. The sample includes 10 companies:**

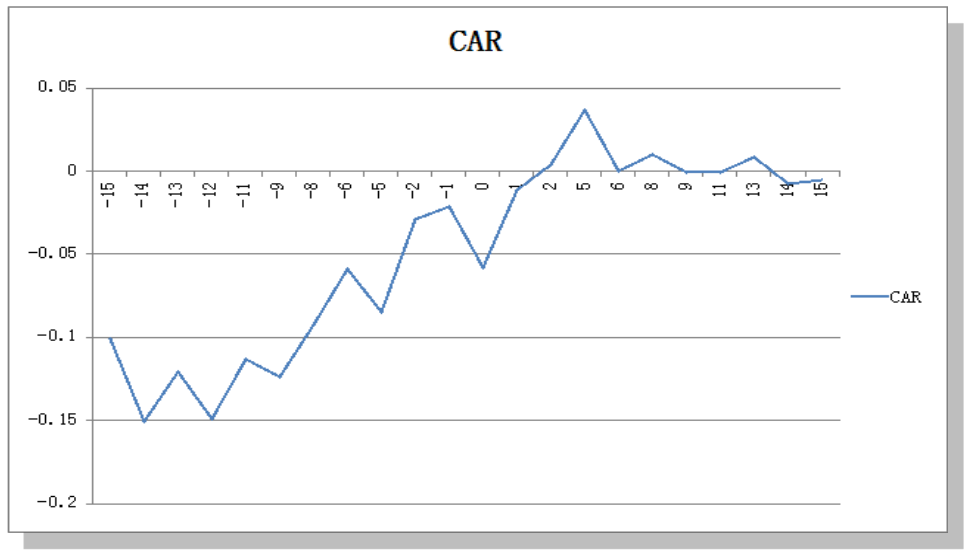

Then, I compared volatilities of stock returns before and after the ADR listing date. The event window used for this period remained the same is [-105;-15] and [15;105] days fort the after period. Variance ratios are computed as follows:

$$R = \frac{\mathrm{var}\, after}{\mathrm{var}\, before} \quad (11),$$

The rates for the four Chinese firms are greater than 1 and were based on F ratio test, three of which are statistically significant. This means that the sample of this firm experienced large fluctuations in the earnings after the listing of ADRs. If the ratio is greater than 1, the variability of stock returns is increased after ADR was introduced, and vice versa. Six of the researched companies has a contrast ratio of less than 1, four of which are statistically significant at 5% confidence level, what means that the variance of the returns of these firms fell on the listing day.

**Table 3-3 The Variance Ratios for the Sample Companies:**

| Company Name | Variance ratio |
|---|---|
| 1  Sinopec Shanghai Petrochemical Co. Ltd. | 2,018020743* |
| 2  Guangshen Railway Co., Ltd. | 2,45247199 |
| 3  China Petroleum &Chemical | 0,886547693* |
| 4  Huaneng Power International | 0,601749763* |
| 5  China Southern Airlines Co. Ltd | 0,770311153 |
| 6  China Life Insurance | 1,967594012* |
| 7  China Eastern Airlines Co. Ltd | 0,781459527* |
| 8  Aluminum Corp. of China Ltd | 1,125571952* |
| 9  Petro China | 0,5502066* |
| 10 China Unicom | 0,7480064 |

The volatility of earnings for the entire sample varied inconsistently and the four-firm experience increased the volatility of local equity returns. The result can be attributed to the low degree of information transparency between the local and US markets.

## *Chapter IV Conclusion*

This paper examined the rate of return of Chinese firms listed on local and overseas markets and cross-listing impact on local economy. Cross-listing is especially common occasion for the companies that started to perform on a small market, but grew into a bigger market. In the second chapter it was found that shares of the same companies when went public on different markets had a different rate of return. A-shares that listed in Chinese market haven't experienced a high volatility performance, comparable with N-shares, listed on US market, of the same Chinese firms. Using the CAPM model, I compared expected rate of return of both kind of shares and revealed that N-shares had a negative expected rate of return, what was caused by turmoil of the economy as a whole. Because of the financial crisis 2007-2008, during which the broader market lost over 50% of its value, the U.S. market was under great pressure and the stock market experienced a high volatility. Therefore listing of Chinese firms on American stock market had a different rate of return with listing in Chinese stock market. Overall findings document that the cross-listing was beneficial for emerging markets, as China, because investors was appealed to invest in less risky environment.

After that, by using a traditional event study method, followed by GARCH process, and by investigating the Chinese market behavior to the listing abroad, I tested the hypothesis of beneficial cross-listing effect on Chinese companies. The findings in the Chapter III indicate that most of them are not significant. Significant negative average abnormal return was found on the day of ADR listing, which is consistent with previous study on beneficial cross-listing effect on companies from another emerging market as Russia (Smirnova 2004). One of possible explanations might be the small sample of researched companies, and the market beta may be overestimated as local stock returns might have been raised after the announcement day and abnormal return may be underestimated. Cumulative abnormal return is helpful for determining how accurate the asset pricing model is in calculating expected return, usually intended to provide analysts with longer-term information about the effects of a major event on a stock's price, and can also serve as a measure of the stock's overall stability, allowing a more accurate assessment of the stock's true worth. In Chapter III, CAR of Chinese firms was climbing from negative to positive after the first listing date, what supports the hypothesis of beneficial cross-listing effect on local economy. However, most of results are statistically insignificant, what suggests that introduction of ADRs in US market does not have a significant impact on the underlying Chinese stocks and consistent with Martell's, Rodriguez' and Webb's research, examining the risk and return of several Latin American equities issues following the introduction of their ADRs in US market.

When considering variance of the returns after the cross listing, the empirical analysis shows that variance of returns for the most researched companies. An increase in variance is connected to more private information acquired by informed traders after the cross listing. High variance associated with a higher risk, but also with a higher return. For most of the researched companies, the variance after the ADR listing provides us with the evidences that more trades were executed on both markets.

The most important needed o be mentioned, that the fall in the rate of the SSE Index does not mean that the return of N shares is a bad thing for the Chinese local market. Decrease in the rate of return shows that the cost of capital is declining. At the market and the New York market, avoiding the marginalization of the A-share market, may also lead to the decline of market risk and investment philosophy.

## 参考文献/*References*

［1］Doidge C. , Karolyi A. , and Stulz R. , 2004 Why are foreign firms listed in the U.S worth more? Journal of Finance Economics, vol. 71, p. 205-238；

[2] Smirnova E. , 2007 Impact of Cross-listing on Local Stock Returns ： Case of Russian ADRs；

[3] 孙会兵 ， 2007 交叉上市的市场效应研究——基于中国A 股市场H回归的证据；

[4]Domowitz, Glen and Madhavan , 1998 International Cross-listing and order flow migration: Evidence from emerging market；

[5] Freedman R. , 1989 A theory of the impact of international Cross-listing；

[6] Jayaraman, Shastri and Tandon , 1993 The impact of international cross-listing on risk and return: The evidence from the ADRs；

[7] Miller D. P. , 1999 The market reaction to international Cross-listing: Evidence from the depositary receipts；

[8] Yarovskaya, 2012 International Cross-listing: Evidence from Russian ADRs；

[9] Gerasymenko J. , 2009 Cross-listing effect and Local stock returns: Evidence from Ukranian ADRs；

[10] Diebold, Lim, Lee (1992). A note on conditional heteroscedasticity in the market model. Journal of Accounting, Auditing, and Finance；

[11] Dodd and Warner (1983)， On corporate governance: A study of proxy contests. Journal of Financial Economics ；

[12] Hamilton (1994), Time Series Analysis. Princeton University Press. Hargis, K.

Evidence from the international cross-listings. Working paper.；

[13] Goldman Sachs and Co.   Hargis and Ramanlal (1998). When does internationalization enhance the development of domestic stock market，Journal of Financial Intermediation；

[14] Gillis, Paul L. "Variable Interest Entities in China," Accounting Matters (Sept. 18, 2012);

[15] Dutton, Edward Drew and Wu, "Investing in China:New Risks?" (2011);

[16] Ball, R., Brown, Ph. (1968), An Empirical Evaluation of Accounting Income Numbers, Journal of Accounting Research, 6 (1968) 159–178;

[17] Korczak, P., Bohl, M.T. (2005), Empirical evidence on cross-listed stocks of Central and Eastern European companies, Emerging Markets Review, 6 (2005) 121–137.；

[18] Mittoo, U.R. (2003), Globalization and the value of US listing: Revisiting Canadian evidence, Journal of Banking & Finance, 27 (2003) 1629–1661；

[19] By Thomas Kenny, （2014, An Explanation of Fed Tapering and its Impact on the Markets

[20] Richard Dobbs and Marc H. Goedhart, (2008), Why cross-listing shares doesn't create value

[21] 数据来源（www.data.worldbank.org. , www.tradingeconomics.com , www.finance.sina.com.cn , www.bloomberg.com ）

## 附录 A/Appendix A:

The Complete List of Chinese ADRs trading on the US Exchanges as of Sept,2014 are listed below:

| S.No. | Company | Ticker | Exchange |
|---|---|---|---|
| 1 | 21Vianet | VNET | NASDAQ |
| 2 | 51job | JOBS | NASDAQ |
| 3 | 58.com | WUBA | NYSE |
| 4 | 500.com | WBAI | NYSE |
| 5 | Acorn | ATV | NYSE |
| 6 | Actions Semiconductor | ACTS | NASDAQ |
| 7 | Agria | GRO | NYSE |
| 7a | Alibaba Group Holding Ltd | BABA | NYSE |
| 8 | Airmedia | AMCN | NASDAQ |
| 9 | Aluminum Corporation of China | ACH | NYSE |
| 10 | ATA | ATAI | NASDAQ |
| 11 | Autohome | ATHM | NYSE |
| 12 | AutoNavi | AMAP | NASDAQ |
| 13 | Baidu | BIDU | NASDAQ |
| 14 | Bitauto | BITA | NYSE |
| 15 | Bona Film | BONA | NASDAQ |
| 16 | Changyou.com | CYOU | NASDAQ |
| 17 | Charm Communications | CHRM | NASDAQ |
| 18 | Cheetah Mobile | CMCM | NYSE |
| 19 | ChinaCache | CCIH | NASDAQ |
| 20 | China Digital TV Holding | STV | NYSE |
| 21 | China Distance Education | DL | NYSE |
| 22 | China Eastern Airlines | CEA | NYSE |
| 23 | China Finance Online | JRJC | NASDAQ |
| 24 | China Life Insurance | LFC | NYSE |
| 25 | China Lodging | HTHT | NASDAQ |
| 26 | China Ming Yang Wind Power Group Ltd | MY | NYSE |
| 27 | China Mobile | CHL | NYSE |
| 28 | China National Offshore Oil-CNOOC | CEO | NYSE |
| 29 | China Nepstar Chain Drugstore | NPD | NYSE |
| 30 | China New Borun | BORN | NYSE |
| 31 | China Petroleum & Chemical | SNP | NYSE |
| 32 | China Southern Airlines | ZNH | NYSE |
| 33 | China Sunergy | CSUN | NASDAQ |
| 34 | China Techfaith Wireless Communication | CNTF | NASDAQ |
| 35 | China Telecom | CHA | NYSE |

| 36 | China Unicom | CHU | NYSE |
| 37 | China Xiniya Fashion | XNY | NYSE |
| 38 | China Zenix Auto International | ZX | NYSE |
| 39 | CNInsure | CISG | NASDAQ |
| 40 | Concord Medical Services | CCM | NYSE |
| 41 | Country Style Cooking Restaurant | CCSC | NYSE |
| 42 | CTrip.com International | CTRP | NASDAQ |
| 43 | Daqo New Energy | DQ | NYSE |
| 44 | E-Commerce China Dangdang | DANG | NYSE |
| 45 | E-House (China) | EJ | NYSE |
| 46 | eLong | LONG | NASDAQ |
| 47 | Guangshen Railway | GSH | NYSE |
| 48 | Hanwha SolarOne | HSOL | NASDAQ |
| 49 | Home Inns & Hotels Management | HMIN | NASDAQ |
| 50 | Huaneng Power International | HNP | NYSE |
| 51 | iDreamSky Technology | DSKY | NASDAQ |
| 52 | IFM Investments | CTC | NYSE |
| 53 | iKang Healthcare Group | KANG | NASDAQ |
| 54 | iSoftStone | ISS | NYSE |
| 55 | JA Solar | JASO | NASDAQ |
| 56 | JD.com | JD | NASDAQ |
| 57 | Jiayuan.com | DATE | NASDAQ |
| 58 | JinkoSolar | JKS | NYSE |
| 59 | Jumei | JMEI | NYSE |
| 60 | Kingtone Wirelessinfo Solution | KONE | NASDAQ |
| 61 | KongZhong | KONG | NASDAQ |
| 62 | Ku6 Media | KUTV | NASDAQ |
| 63 | LDK Solar | LDK | NYSE |
| 64 | Le Gaga Holdings | GAGA | NASDAQ |
| 65 | Leju Holdings | LEJU | NYSE |
| 66 | Lentuo International | LAS | NYSE |
| 67 | LightInTheBox | LITB | NYSE |
| 68 | Mecox Lane | MCOX | NASDAQ |
| 69 | Mindray Medical International | MR | NYSE |
| 70 | NetEase | NTES | NASDAQ |
| 71 | New Oriental Education & Technology | EDU | NYSE |
| 72 | Noah Holdings | NOAH | NYSE |
| 73 | NQ Mobile | NQ | NYSE |
| 74 | Ossen Innovation | OSN | NASDAQ |
| 75 | Perfect World | PWRD | NASDAQ |
| 76 | PetroChina | PTR | NYSE |
| 77 | Phoenix New Media | FENG | NYSE |
| 78 | Qihoo 360 Technology | QIHU | NYSE |
| 79 | Qunar | QUNR | NASDAQ |

| 80 | Rda Microelectronics | RDA | NASDAQ |
| 81 | ReneSola | SOL | NYSE |
| 82 | Renren | RENN | NYSE |
| 83 | Shanda Games | GAME | NASDAQ |
| 84 | Sinopec Shanghai Petrochemical | SHI | NYSE |
| 85 | Sky mobi | MOBI | NASDAQ |
| 86 | SouFun | SFUN | NYSE |
| 87 | Sungy Mobile | GOMO | NASDAQ |
| 88 | TAL Education | XRS | NYSE |
| 89 | Taomee Holdings | TAOM | NYSE |
| 90 | The9 | NCTY | NASDAQ |
| 91 | Trina Solar | TSL | NYSE |
| 92 | Vimicro International | VIMC | NASDAQ |
| 93 | Vipshop | VIPS | NYSE |
| 94 | VisionChina Media | VISN | NASDAQ |
| 95 | Weibo Corporation | WB | NASSDAQ |
| 96 | WSP Holdings | WH | NYSE |
| 97 | WuXi Pharmatech | WX | NYSE |
| 98 | Xinyuan Real Estate | XIN | NYSE |
| 99 | Xueda Education | XUE | NYSE |
| 100 | Xunlei | XNET | NASDAQ |
| 101 | Yanzhou Coal Mining | YZC | NYSE |
| 102 | Yingli Green Energy | YGE | NYSE |
| 103 | Youku.com | YOKU | NYSE |
| 104 | YY Inc. | YY | NASDAQ |
| 105 | Zhaopin LTD | ZPIN | NYSE |
| 106 | Zuoan Fashion | ZA | NYSE |

## 附录 B/Appendix B:

| Code | $R_{const}$ | $R_{sse}$ | $R_{nyse}$ | #lags arch(); garch() | $w_i$ |
|------|-------------|-----------|------------|-----------------------|-------|
| YZC | -0,014979 | 0,010893 | 1,434841 | 1;1 | 0,009228081 |
| GSH | 0,004473 | 0,27408 | 1,73238 | 1;1 | 0,009737719 |
| SNP | -0,000108 | 0,904283 | 0,505306 | 1;1 | 0,190677272 |
| HNP | 0,004582 | 0,062199 | -0,074935 | 1;1 | 0,032999035 |
| ZHN | 0,00099 | -0,042267 | 0,103376 | 1;1 | 0,016435813 |
| LFC | 0,005072 | -0,3025 | 0,599311 | 1;1 | 0,258076846 |
| ACH | 0,002909 | 0,024156 | 2,524365 | 1;1 | 0,015798766 |
| PTR | 0,002563 | -0,148161 | 1,30882 | 1;1 | 0,437614896 |
| CEA | -1.03E-18 | -7.59E-53 | 1 | 1;1 | 0.02113176 |